\documentclass[12pt,english,dvips]{scrartcl}
\usepackage{amssymb}

\begin{document}


\begin{titlepage} \vspace{0.2in} 

\begin{center} {\LARGE \bf 
Quasi-Isotropization of the Inhomogeneous Mixmaster Universe 
Induced by an Inflationary Process
\\} \vspace*{0.8cm}
{\bf A. A. Kirillov${}^*$ and G. Montani${}^{**}$ 
}\\ \vspace*{1cm}
${}^*$)Institute for Applied Mathematics and Cybernetics\\ 
10 Ulyanova str., Nizhny Novgorod, 603005, Russia\\ 
\vspace*{0.8cm}
${}^{**}$)ICRA---International Center for Relativistic Astrophysics\\ 
Dipartimento di Fisica (G9), 
Universit\`a  di Roma, ``La Sapienza", 
Piazzale Aldo Moro 5, 00185 Rome, Italy
\vspace*{1.8cm}

 \vspace*{1cm} \\ 

{\bf   Abstract  \\ } \end{center} \indent
We derive a ``generic'' inhomogeneous ``bridge'' solution for a cosmological model 
in the presence of a real self-interacting scalar field. This solution connects 
a Kasner-like regime to an inflationary stage of evolution 
and therefore provides a dynamical mechanism 
for the quasi-isotropization of the universe. 
In the framework of a standard Arnowitt-Deser-Misner Hamiltonian 
formulation of the dynamics and by adopting 
Misner-Chitr\`e-like variables, 
we integrate the Einstein-Hamilton-Jacobi equation corresponding 
to a ``generic'' inhomogeneous cosmological model 
whose evolution is influenced by the coupling with a 
bosonic field, expected to be
responsible for a spontaneous symmetry breaking configuration.
The dependence of the detailed evolution of the universe on the 
initial conditions is then appropriately characterized. 

\end{titlepage}

\section{Introduction}

As is well known \cite{BKL70,BKL82} (see also \cite{K93}-\cite{M95}) the general
solution of the Einstein equations near a cosmological singularity exhibits
an oscillatory stochastic behavior. This feature of the very early universe
is in striking contrast with the universe as described by the well-tested
theory of the standard cosmological model \cite{KT90}, which is based on the
highly symmetric Friedmann-Robertson-Walker geometry. However, the
experimental evidence for the homogeneous and isotropic character of our
actual universe concerns relatively late stages of evolution. Indeed the
good agreement of the light element nucleosynthesis prediction with the
observed abundances implies that the standard cosmological model is surely
valid after $10^{-3}$-$10^{-2}$ sec from the big bang, but says nothing
about the very early dynamics before this time.

In this respect, by observing that the Friedmann-Robertson-Walker metric is
an unstable solution of the Einstein equations when regarded as running
backwards in time \cite{LK63}, then from the existence of structures in the
universe \cite{MB95} like galaxies and clusters of galaxies, we may infer
(even in the presence of an inflationary scenario) that such symmetric
geometry cannot continue all the way up to the initial singularity. In fact,
the clumpiness of the universe indicates the necessity for very early
perturbations of its homogeneity and isotropy, which unavoidably ``explode''
when approaching the big bang\footnote{
Actually, this represents a rather weak point since the universe could start out 
in a completely homogeneous and isotropic state, while primeval perturbations which
necessary for the origin of structures might appear later during the 
inflationary era.
}.

The instability of the FRW metric,
when regarded backward in time, means that
there exists some moment $t_*$ before which
the evolution of the Universe was
to be described by a ``generic`` inhomogeneous model,
or , by other words,
when the BKL (Belinskii - Khalatnikov - Lifshitz)
picture \cite{BKL70,BKL82} holds.
In general this moment
represents a free parameter which depends
on initial conditions and, in particular,
on specific properties of matter.
In vacuum case the applicability of the
BKL picture was shown \cite{K93} to be described
by the inequality $L_{h}\ll L_{in}$
where $L_{h}\sim t$ is the horizon size and $L_{in}$
is the characteristic scale of
inhomogeneity (or, to be more precise,
the mean geometrical value of all leading
inhomogeneity scales
\footnote{We recall that inhomogeneity
of the metric is described by a set of distinct length scales
$L^A_{in}$ which correspond to different
terms in the spatial scalar curvature.
The anisotropic evolution leads to the increase of
some curvature terms
(i.e., the corresponding scales decrease with
respect to the horizon size) which
causes the transition between Kasner regimes
as described in Ref.~\cite{BKL82}. From the
qualitative standpoint, the transition of a
Kasner epoch occurs when the
horizon size matches the smallest of the scales
$L_{h}\sim L^A_{in}$
and therefore the above inequality is violated
at least for one of the scales.
However, the duration of such violations are
small compared with the duration
of Kasner epochs and for the average geometrical
value of the scales, this inequality still holds. 
Anyway the greater the inequality,
the better the BKL picture works.})

Thus, in the vacuum case the moment $t_*=t_{in}$
when the Mixmaster phase
(i.e. the oscillatory regime) 
ends 
corresponds to the situation
$L_{h}\sim L_{in}$,
which can be roughly considered as a
boundary of the BKL approximation
(in the early stages
only the stiff matter influences the
evolution of the metric).
The reversibility of the
Einstein equations
means that the BKL picture works in
both directions of time
(which is explicitly realized in
the so-called billiard representation \cite{K93,KM95}).
The fact that $t_{in}$ is a free parameter
means that near the singularity at a moment $t_0$,
the initial conditions can always be
chosen in such a way that $t_0\ll t_{in}$ and there is a
defined period (which depends on the degree of inhomogeneity
of the metric) when the BKL
picture still works even with the increase of time.

We note that there exists another important
moment $t_m$ when the matter ``switches on"
(i.e., starts to influence the
evolution of metric).
This moment $t_m$ represents another free parameter
specified by initial conditions for matter.
Therefore, the actual moment $t_*$ when
the Mixmaster evolution (BKL picture) breaks down depends
on the relation between
this two parameters $t_m$ and $t_{in}$.

In the present paper we consider a restricted
region of initial conditions
in which the inequality $t_m\ll t_{in}$ holds, which
means that the matter
starts to dominate deep inside the BKL
regime. From the physical standpoint
this restriction means that the transition from
the one BKL regime
to another
takes place when locally, i.e., on
the scale of causal connection, the
universe still looks like a homogeneous model,
for the inequality
$L_h\sim t\ll t_{in}$ is fulfilled,
while global properties are
described by a general inhomogeneous metric. 
Therefore, it is natural to expect that
some results obtained for
homogeneous models can also be applied there.
In particular, if $t_m$
corresponds to the beginning of an inflationary
period, then the inequality
$L_h\ll t_{in}$ remains valid from the BKL era
through the inflationary period
\footnote{We recall that during the inflation
the inhomogeneity scales increase more
rapidly than does the horizon size.} 
and therefore we can use the results
of Ref.~\cite{W83}.

The chaotic nature of the evolution (both temporally and spatially) implies
that the geometry of the very early universe should be described by a
stationary statistical distribution \cite{K93} (see also \cite{B82} -\cite
{IM01}). Indeed in this context we may speak about geometry only in an
average sense; it turns out that mean values of all geometrical quantities
(lengths, scalar products, etc.) during the oscillatory regime are unstable
(higher moments have the same order of magnitude as the average values) and
therefore near the singularity the universe does not possess a stable
background. We remark that the same situation holds in the quantum evolution
of the inhomogeneous Mixmaster \cite{M69q,K97} universe, although in the
quantum case the statistical distribution has a different (but somehow
related) nature.

These considerations first pose the problem of the origin of a stable
background and second how this background could arise out of this chaos 
compatible with
the notion of isotropy (on the basis of any acceptable early history of our
actual universe). However, the strong correlation between the appearance of
a stable background and its isotropic character is a key feature of the very
early cosmology. Either on a quantum level or on a classical one the
isotropic component of the metric tensor (i.e. the volume of the universe)
is a monotonic function of the time variable (which may actually be taken as
the time coordinate itself) and therefore does not contain any physical
degrees of freedom, which are instead entirely contained in the anisotropic
components. In other words a stable background metric can only appear when
the anisotropy of the universe is sufficiently suppressed \cite{K93,K97}.

In vacuum inhomogeneous models this problem was considered first in \cite
{KMq97}, which outlines how a classical background can arise from the Planckian
epoch of the universe essentially when the oscillatory regime is over, i.e.
at the moment when the characteristic scale of inhomogeneity $L_{in}$
matches the horizon size $L_{h}$. Analogously, in the presence of matter
described by vector fields, the background appears when the horizon size
reaches the minimal scale between the one related to the vector field and
the characteristic scale of inhomogeneity \cite{KS} (for a discussion of
chaos in superstring cosmology relative to Einstein-dilaton-$p$-form fields
see \cite{DH} and \cite{DH1}). It is important that in both cases the
anisotropy of the universe decays (i.e. it becomes smaller and smaller)
during the natural Mixmaster-like evolution from the initial singularity (it
may be worth noting the analysis presented in \cite{A} on quiescent
cosmological singularities).

In this paper we consider the origin of a background space when a real
self-interacting scalar field is present in the universe. As we shall see in
this case the appearance of a background depends on the initial conditions
(to be assigned on a nonsingular space-like hypersurface); in the
configuration (phase) space there are two regions of initial conditions for
which the evolution behaves in qualitatively different ways.

The first region corresponds to the case when the potential term of the
scalar field becomes a dominating term before the end of the Mixmaster
evolution ($L_{h}\sim L_{in}$) (e.g., such a region can be characterized by
the inequality $L_{c}\ll L_{in}$, $L_{c}$ denoting the Compton length
associated with the scalar field, so that the horizon size matches first the
Compton scale). In this case the scalar field through its energy completely
governs the quasi-isotropization process (i.e. the process which gives
origin to a stable background). The appropriate region of initial conditions
contains a subregion which corresponds to an inflationary-like evolution of
the universe. The second region corresponds to the case when the scalar
field potential remains small and, from a qualitative point of view, the
origin of a stable background occurs in the same way as in vacuum models.

Below we will consider the first region only, having in mind the idea that a
classical quasi-isotropic universe may emerge, up to suitable initial
conditions, from general cosmological dynamics, essentially by virtue of an
inflationary expansion due to the potential term of the real scalar field.
Indeed the main result of this paper is to show the existence of a set of
initial conditions of a nonzero measure, corresponding to which the
anisotropy of the universe decays exponentially during an inflationary phase
(in homogeneous models the inflationary phase and the isotropization of the
Universe has been considered in Ref.~\cite{W83}). Thus the analysis of the
``generic'' cosmological solution shows how the inflation phenomenon is the
``bridge'' between the chaoticity near the big bang (indeed in the presence
of a real scalar field the Mixmaster contains only a finite number of
oscillations \cite{BK73,B99}) and the phenomenology of the standard
cosmological model.

In section 2 we develop the standard Arnowitt-Deser-Misner (ADM) Hamiltonian
formulation \cite{ADM,Grav} which is at the foundation of our derivation in
section 3 of a ``generic'' solution of the Einstein-Hamilton-Jacobi equation
in presence of a real self-interacting scalar field. Such a solution
interpolates between a Kasner-like regime and an inflationary scenario and
is to be regarded as the main result of this paper (for a related discussion
of the Bianchi I model in the path integral formalism see \cite{B85}). In
Section 4 we provide a reformulation of the system dynamics in terms of
Misner-Chitr\`e-like variables, in order to give the most appropriate
framework for the analysis presented in Section $5$ and devoted to emphasize
the modification induced in the details of the universe evolution by
assigning different initial conditions to the dynamical quantities involved
in the problem. Thus this analysis defines the range of existence for the
solution we have obtained.

\section{Hamiltonian Formulation of the Dynamics}

Let us start by fixing the dynamical framework for our investigation of the
``generic'' inhomogeneous dynamics. We first observe that the dynamical
regime we find will be regarded as ``generic'' in the sense that it
possesses the number of physically arbitrary functions of the spatial
coordinates (i.e. real degrees of freedom of the physical system) required
to specify a generic Cauchy problem on a nonsingular spatial hypersurface
(having in mind one which is arbitrarily close to the big bang) for a
generic $(n + 1)$-dimensional space-time, containing a real self-interacting
scalar field, the number of physically independent degrees of freedom is $%
n(n - 1)$. Indeed this number is $(n + 1)(n - 2)/2$ for the gravitational
field, i.e. the number of independent polarizations of a gravitational wave,
plus 1 for the real scalar field, but both these fields satisfy second order
equations.

The line element of a generic $(n + 1)$-dimensional space-time (for the sake
of generality we will consider the most general case and the results for our
actual universe will follow immediately by setting $n = 3$) admits the
following standard (ADM) representation:

\begin{equation}
ds^{2}=N^{2}dt^{2}-g_{\alpha \beta } ( dx^{\alpha }+N^{\alpha }dt) (
dx^{\beta }+N^{\beta }dt) .  \label{met}
\end{equation}
Then the Einstein-Hilbert action takes the form (in what follows we use
units of the Planck length)
\begin{equation}
I=\int d^{n}xdt\left\{ \pi ^{\alpha \beta }\frac{\partial }{\partial t}
g_{\alpha \beta }+\Pi _{\phi }{\frac{\partial \phi }{\partial t}}
-NH^{0}-N^{\alpha }H_{\alpha }\right\} ,  \label{act}
\end{equation}
where the super-Hamiltonian $H^{0}$ and the super-momentum $H_{\alpha }$ are
respectively
\begin{equation}
H^{0}=\frac{1}{\sqrt{g}}\left\{ \pi _{\beta }^{\alpha }\pi _{\alpha }^{\beta
}-\frac{1}{n-1}\left( \pi _{\alpha }^{\alpha }\right) ^{2}+g\left(
W-R\right) \right\} ,  \label{hamcn}
\end{equation}
\begin{equation}
H_{\alpha }=-2\nabla _{\beta }\pi _{\alpha }^{\beta }+\Pi _{\phi }\partial
_{\alpha }\phi ,  \label{momcn}
\end{equation}
where $W(\phi )={\frac{1}{2}}\left\{ g^{\alpha \beta }\partial _{\alpha
}\phi \partial _{\beta }\phi +V(\phi )\right\} $ and $R$ is the spatial
scalar curvature constructed from the spatial metric $g_{\alpha \beta }$.

A fundamental step in our investigation consists of rewriting the above
(ADM) formulation in a form which is useful for our purposes, but which
retains its degree of generality. It turns out to be convenient to use the
so-called generalized Kasner-like parameterization of the dynamical
variables in terms of $n$ logarithmic scale variables $q_{a}$ and $n$
spatial frame covector fields $\ell _{\alpha }^{a}$ dual to a spatial frame
consisting of $n$ vectors $L_{a}^{\alpha }$ (having inverse component
matrices). The $n$-dimensional metric components and their conjugate momenta
are represented in terms of these variables as follows

\begin{equation}
g_{\alpha \beta }=\sum_{a}\exp \left\{ q^{a}\right\} \ell _{\alpha }^{a}\ell
_{\beta }^{a}\,,\qquad \pi _{\beta }^{\alpha }=\sum_{a}p_{a}L_{a}^{\alpha
}\ell _{\beta }^{a}\,,  \label{ksnpr1}
\end{equation}
where the covectors $\ell _{\alpha }^{a}$ can contain only $n(n-1)$
arbitrary functions of the spatial coordinates.  By definition the
Kasner vectors are eigenvectors for both the momentum matrix $\pi _{\beta
}^{\alpha }$ and the metric $g_{\alpha \beta }$ and therefore in the case
of a generic point in spacetime, such a decomposition is unique ($n\left( n-1\right) $
arbitrary functions contained in Kasner vectors $\ell _{\alpha }^{b}$ and $%
2n $ functions $q^{a}$ and $p_{a}$ replace the $n\left( n+1\right) $
functions contained in $g_{\alpha \beta }$ and $\pi _{\beta }^{\alpha }$).
This general form of the metric is the most suitable for treating the
Kasner-like regime.

A further refinement of the parameterization can be made by separating the
different types of contributions to the matrix $\ell _{\alpha }^{a}$ as
follows (e.g., see Ref.~\cite{BKL82}):
\begin{equation}
\ell _{\alpha }^{a}=U^{a}{}_{b}S_{\alpha }^{b},\qquad U^{a}{}_{b}\in
SO(n),\qquad S^{a}{}_{\alpha }=\delta ^{a}{}_{\alpha }+R^{a}{}_{\alpha }
\label{ksnpr3}
\end{equation}
with $R^{a}{}_{\alpha }$ denoting a triangle matrix ($R^{a}{}_{\alpha }=0$
if $a<\alpha $) and, therefore, it contains\ only $n(n-1)/2$ arbitrary
functions of coordinates, while the rest $n(n-1)/2$ arbitrary functions are
included in the rotation matrix $U_{b}^{a}$. By substituting (\ref{ksnpr1})
and (\ref{ksnpr3}) into (\ref{act}), we rewrite the action functional in the
form (any repeated index is to be regarded as summed) 
\begin{equation}
I=\int (p_{a}{\frac{\partial q^{a}}{\partial t}}+T_{a}^{\alpha }{\frac{%
\partial R_{\alpha }^{a}}{\partial t}}+\Pi _{\phi }{\frac{\partial \phi }{%
\partial t}}-NH^{0}-N_{\alpha }H^{\alpha })d^{n}xdt,  \label{actksnr}
\end{equation}
where $T_{a}^{\alpha }=2\sum_{b}p_{b}L_{b}^{\alpha }U_{a}^{b}$ and the
Hamiltonian and the momentum constraints take the form
\begin{equation}
H^{0}={\frac{1}{\sqrt{g}}}\left\{ \sum_{a}p_{a}^{2}-{\frac{1}{n-1}}%
(\sum_{a}p_{a})^{2}+{\frac{1}{2}}\Pi _{\phi }^{2}+U\right\} ,  \label{hcnstr}
\end{equation}
\begin{equation}
H_{\alpha }=-{\frac{1}{\sqrt{g}}}\partial _{\beta }\left( \sqrt{g}%
T_{a}^{\beta }S^{a}{}_{\alpha }\right) +p_{a}\partial _{\alpha
}q^{a}+T_{a}^{\beta }\partial _{\alpha }R_{\beta }^{a}{+}\Pi _{\phi
}\partial _{\alpha }\phi .  \label{sm}
\end{equation}
We note that due to the property $U_{ij}\partial _{t}U_{ik}{=-}%
U_{ik}\partial _{t}U_{ij}$ the time derivative of the matrix $U^{a}{}_{b}$
drops out from the expression (\ref{actksnr}) 
and besides, only $n\left(
n-1\right) /2$ components of the matrix $T_{a}^{\alpha }$ (i.e., with $%
a>\alpha $) should be considered as independent functions which are \
canonically conjugate to the triangle matrix variables $R^{a}{}_{\alpha }$.

In the super-Hamiltonian $H^{0}$ constraint the quantities $R_{\alpha }^{a}$
and $T_{a}^{\alpha }$ contribute only to the spatial curvature in the term $%
U=g\left( W-R\right) $ and for the case of $n=3$ the functions $R_{\alpha
}^{a}$ are connected purely with transformations of the coordinate system
and may be removed by solving the super-momentum constraints $H^{\alpha }=0$ 
\cite{K93} (which expresses independent components of $T_{a}^{\alpha }$ as
functions of $p_{a}$, $q^{a}$, $\Pi _{\phi }$, and $\phi $). In the
multidimensional case, however, the functions $R_{\alpha }^{a}$ contain ${%
\frac{n(n-3)}{2}}$ dynamical functions as well which cannot be removed by
coordinate transformations. However, in what follows we shall use model
representations for the potential term $U$ in which the dependence of $U$ on 
$R_{\alpha }^{a}$ and $T_{a}^{\alpha }$ can be neglected (e.g., the
generalized Kasner model (GKM) corresponds to the case where we neglect the
spatial curvature, the inhomogeneous Mixmaster model (IMM) corresponds to
the case where we replace the spatial curvature term with a set of infinite
potential walls). Therefore in these models the super-Hamiltonian will not
depend explicitly on $R_{\alpha }^{a}$ and $T_{a}^{\alpha }$ and these
functions will have a passive character and can be considered separately.
Indeed, in the case of the GKM or IMM the evolution of these functions is
completely governed by the supermomentum constraint (\ref{sm}) which can be
used to express $n$ independent functions among $R_{\alpha }^{a}$ and $%
T_{a}^{\alpha }$ via the rest passive ($R_{\alpha }^{a}$ and $T_{a}^{\alpha
} $) and dynamical ($p_{a}$, $q^{a}$, $\Pi _{\phi }$, $\phi $) functions. We
note that in the GKM all the passive functions represent merely constants of
the motion (e.g., see Ref.~\cite{LK63}), while in the IMM the oscillatory
evolution is accompanied by a rotation of the Kasner vectors which is
completely determined by the momentum constraint (e.g., see Ref.~\cite{BKL82,BK73}
).

After having shown how the above formal decomposition of the metric
variables into scale functions and ``reference'' vectors acquires a precise
dynamical meaning in the above action, we must make a key distinction
among the scale functions themselves by extracting their isotropic component
from the anisotropic ones. This is accomplished by further refining the
metric parametrization by introducing coordinates on the space of scale
variables which are quasi-orthonormal with respect to the DeWitt
minisuperspace metric following Misner \cite{KM95} 
\begin{equation}
q^{a}=A^{a}{}_{j}{\beta }^{j}+\alpha \,\quad {\beta }^{n}=[n(n-1)/2]^{-1/2}%
\phi \,,  \label{harm}
\end{equation}
where $j=1,...,n-1$, and the suitably chosen constant matrix $A_{j}^{a}$
obeys the conditions 
\begin{equation}
\sum_{a}A^{a}{}_{j}=0,\qquad \sum_{a}A^{a}{}_{j}A^{a}{}_{k}=n(n-1)\delta
_{jk}\,,  \label{mA}
\end{equation}
for example 
\begin{equation}
A^{a}{}_{j}=\sqrt{{\frac{n(n-1)}{j(j+1)}}}(\theta ^{a}{}_{j}-j\delta
^{a}{}_{j}),\qquad \theta ^{a}{}_{j}=\left\{ 
\begin{array}{ll}
1,\quad & j>a\,, \\ 
0,\quad & j\leq a\,.
\end{array}
\right.  \label{matrA}
\end{equation}
Since $g=\exp \left( n\alpha \right) $, we see that $\alpha $ corresponds to
the volume or isotropy degree of freedom, while ${\beta }^{j}$ ($%
j=1,2,...,n-1$) describe the anisotropy of the model. When these
variables are sufficiently suppressed in the sense that they asymptotically
approach constant values, we may speak of quasi-isotropization of the model.

Then the action expressed in these variables formally resembles the action
of a relativistic particle moving in a potential (here the index $r$
runs from $1$ to $n$)\newline
\begin{equation}
I=\int \left( P_{r}{\frac{\partial }{\partial t}}{\beta }^{r}+P_{\alpha }{%
\frac{\partial }{\partial t}}{\alpha }-\frac{N}{n\left( n-1\right) \sqrt{g}}%
\left( \sum_{r}P_{r}^{2}+\widetilde{U}-P_{\alpha }^{2}\right) \right)
d^{n}xdt,  \label{eq:redact2}
\end{equation}
where the potential term $\widetilde{U}=n\left( n-1\right) U$ may be viewed
as a ``mass term'' for the particles which depends on dynamical variables
(on the position in the phase space) and is not everywhere positive.

It is important to emphasize that since our goal is to consider simple models (GKM and
IMM), for the sake of simplicity we removed from the above action the
passive functions (the terms $T_{a}^{\alpha }\partial R_{\alpha
}^{a}/\partial t$) as well as the super-momentum terms which as
explained above can be considered separately.

\section{Construction of the Inhomogeneous ``Bridge'' Model}

In this section we derive a ``generic'' inhomogeneous solution connecting
the Kasner-like behavior (to be regarded as one of the Kasner epochs in the
oscillatory regime) with an inflationary regime \cite{Star83}. It is worth
noting that, although our analysis is done on a purely classical level and
in the absence of ultrarelativistic matter, nevertheless from the
qualitative point of view it has a predictive character even in a more
general context. Indeed on the one hand we may expect that during the
last Kasner epoch of the oscillatory regime the so-called {\it quantum
potential}\footnote{
If we separate the wave functional of the universe into its modulus and its
phase, then the latter will satisfy the Hamilton-Jacobi equation
containing an additional ``quantum potential''.} plays, in the
Hamilton-Jacobi equation, the role of a small correction to the classical
potential (which can be modelled by a set of infinite potential walls, e.g.
see the next section). From the other hand, either during the Kasner regime
in the presence of a scalar field or during the inflationary phase, the
contribution of the ultrarelativistic matter is negligible in comparison
with the kinetic \cite{M00} and/or the potential \cite{G81} energy of the
scalar field respectively.

The inflationary solution can be obtained from the action (\ref{eq:redact2})
if we impose restrictions of the form
\begin{equation}\label{rr}
\frac{1}{g}\widetilde{U}\simeq V\left( \phi \right) \simeq const\gg
R\;,\;g^{\alpha \beta }\partial _{\alpha }\phi \partial _{\beta }\phi
\end{equation}
which can be realized by an appropriate process of spontaneous symmetry
breaking as described in the standard literature on this subject (see \cite
{G81}, \cite{L82,L83} and \cite{ST87}). The resulting model given below
describes the inflationary expansion of an inhomogeneous universe in the
limit $g\rightarrow \infty $ (i.e. $\alpha \rightarrow \infty $), while the
Kasner-like regime \cite{LK63} is obtained asymptotically approaching the
singularity for $g\rightarrow 0$ (i.e. $\alpha \rightarrow -\infty $). To
derive such a solution we use the Einstein-Hamilton-Jacobi method since in
the sense mentioned above this theory is the quasiclassical approximation to
quantum gravity and also because it is computationally convenient.

Let us consider the situation where in (\ref{eq:redact2}) the potential
(mass) term can be approximated as 
\begin{equation}
\widetilde{U}=n\left( n-1\right) g\Lambda ,
\end{equation}
where $\Lambda = \Lambda (x^i) \approx const$. These conditions impose
peculiar restrictions on the degree of inhomogeneity of the scalar and
gravitational fields and on the potential form of $V(\phi )$. Then the
Einstein-Hamilton-Jacobi equations are 
\begin{equation}
P_{r} = {\frac{\delta I}{\delta {\beta }^{r}}} , \, P_{\alpha } = {\frac{
\delta I}{\delta \alpha }} , \, \,\,H^{0}(\alpha , {\beta }^{r}, P_{\alpha
}, P_{r}, \Lambda )=0  \label{s}
\end{equation}
or explicitly
\begin{equation}
\sum_{r} \left( {\frac{\delta I}{\delta {\beta }^{r}}} \right) ^2 - \left( {
\frac{\delta I}{\delta \alpha }}\right)^2 + n\left( n-1\right) \exp \left(
n\alpha \right)\Lambda = 0 .  \label{hjs}
\end{equation}

The evolution of the residual variables remaining in the Kasner reference
vectors $\ell ^{a}{}_{\alpha }$ are determined by the super-momentum
constraint equations $H_{\alpha }=0$ and can be expressed via the functions $
(P_{r},{\beta }^{r})$.

The solution of eq.~(\ref{s}) can be expressed in the form 
\begin{equation}
I({\beta }^{r}, \alpha )=\int_{S}\left\{ K_{r} {\beta }^{r}+({\frac{2}{n}K}
_{\alpha }+{\frac{K}{n}} \ln \mid {\frac{K_{\alpha }-K}{K_{\alpha }+K}}\mid
\right\} d^{n}x,  \label{a}
\end{equation}
where $K_{\alpha }(K_r, \alpha )=\pm \sqrt{ \sum_{r}K_{r}^{2}+n\left(
n-1\right) \Lambda \exp \left( n\alpha \right) }$, $K= \sqrt{
\sum_{r}K_{r}^{2}}$ and $K_{r}$ are arbitrary ``constant''  functions of 
the spatial coordinates (i.e. independent
of the time variable). Here $S$ denotes
the whole available spatial domain in which restrictions (\ref{rr}) are fulfilled.

The signs $\pm $ before the square root correspond to the two possibilities
for the variation of the local spatial volume in $S$ (it depends on whether
collapse or expansion of $S$ is considered). Indeed, from (\ref{eq:redact2})
we find that the variation of $\alpha $ is determined by the Hamilton
equation 
\begin{equation}
{\frac{\partial \alpha }{\partial t}}=-{\frac{2NP_{\alpha }}{n\left(
n-1\right) \exp \left( n\alpha /2\right) }}
\end{equation}
and $N/\exp \left( n\alpha /2\right) >0$. In order to further simplify our
analysis, we choose as a time variable the quantity $\alpha $ (i. e, $\frac{
\partial }{\partial t}\alpha =1$), which implies the time gauge
condition $N=n\left( n-1\right) \exp \left( n\alpha /2\right) /\left(
-2P_{\alpha }\right) ${)} (since the lapse function should be positive by
definition, we must have $P_{\alpha }<0$).

Now according to the Hamilton-Jacobi method, we differentiate with
respect to the quantities $K^{r}$ and then by putting the results  equal
to arbitrary ``constant'' functions, we find the solutions describing the
trajectories of the system (${\frac{\delta I}{\delta K^{r}}}={\beta }%
_{0}^{r} $) 
\begin{equation}
{\beta }^{r}(\alpha ,x^{i})={\beta }_{0}^{r}(x^{i})+{\frac{K_{r}}{n\left|
K\right| }}\ln \left| {\frac{K_{\alpha }-K}{K_{\alpha }+K}}\right| \;,
\label{sol}
\end{equation}
where ${\beta }_{0}^{r}(x^{i})$ are arbitrary ``constant'' functions. In the
asymptotic limit $g\rightarrow \infty $ (i.e. $\alpha \rightarrow \infty
\;\Rightarrow \;K_{\alpha }\rightarrow \infty $) the solution (\ref{sol})
transforms into the inflationary solution obtained in \cite{Star83},
corresponding to the quasi-isotropization of the model since the parametric
functions ${\beta }^{r}$ approach (exponentially) the ``constant'' values ${%
\beta }_{0}^{r}(x^{i})$. In the opposite limit $g\rightarrow 0$ (i.e. $%
\alpha \rightarrow -\infty \;\Rightarrow \;K_{\alpha }\rightarrow K$) (\ref
{sol}) transforms into the generalized Kasner solution as it should,
modified by the presence of the scalar field (see the \cite{BK73}) 
\begin{equation}
{\beta }^{r}(\alpha ,x^{i})={\beta }_{0}^{r}(x^{i})-{\frac{K_{r}}{K}}(\alpha
-\alpha _{0})\ ,  \label{k}
\end{equation}
where $\alpha _{0}$ renames the remaining constant terms.

We conclude our analysis by establishing the relation between our time
variable $\alpha $ and the synchronous time $T$, which has a precise
cosmological interpretation. These two variables are connected by the
differential expression 
\begin{equation}
dt = Nd\alpha = -{\frac{n\left( n-1\right) \exp \left( n\alpha /2\right) }{
2P_{\alpha }})} d\alpha .  \label{ggg}
\end{equation}

Now it is easy to see that from the Hamilton equation obtained by varying
the action with respect to $\alpha $ we get, having fixed our time gauge
(and remembering that $H^0 = 0$) the following asymptotic behaviors: 
\begin{equation}
\alpha \rightarrow \infty \; : \; P_{\alpha } \propto -\sqrt{\Lambda }\exp
\left( n\alpha /2\right) \ , \quad \alpha \rightarrow -\infty \; : \;
P_{\alpha } \propto const < 0 .  \label{ggg1}
\end{equation}
By substituting these relations into (\ref{ggg}), we find the expected
(familiar) asymptotic relations (which make evident the character of the two
``opposite'' regimes) 
\begin{equation}
\alpha \rightarrow \infty \; : \; \alpha \propto \sqrt{\Lambda }t
\Rightarrow \sqrt{g}\propto \exp \left( C_1\sqrt{\Lambda }t\right) \, \quad
\alpha \rightarrow -\infty \; : \; \alpha \propto \frac{2}{n}\ln C_2t
\Rightarrow \sqrt{g}\propto t  \label{ggg2}
\end{equation}
where $C_1$ and $C_2$ denote two constant values.

The existence of this solution shows how the inflationary scenario can
provide the necessary dynamical ``bridge'' between the fully anisotropic and
the quasi-isotropic stage of the universe evolution.

\section{Misner-Chitr\`e-like approach}

Though our solution is perfectly characterized by the above Misner-like
variables, nevertheless to make precise the restrictions to be imposed on
the initial conditions for the existence of such an interpolating regime, it
is necessary to investigate a bit in detail the finite oscillating evolution
to the singularity and therefore it is much more convenient to make use of
the so-called Misner-Chitr\`{e}-like variables. \cite{KM95} In order to
introduce these variables, the scale functions $q^{a}$ may instead be
parameterized as follows (see Ref.~\cite{KM97}) 
\begin{equation}
q^{a}=\ln R_{0}^{2}+M_{a}\ln g;\,\,\sum_{a}M_{a}=1\;,\quad a=1,2,...,n\,
\label{SF}
\end{equation}
where we distinguished a slowly varying function of time $R_{0}$, which
characterizes the absolute value of amplitude of the metric functions \cite
{M69,Grav} and is specified by initial conditions (see below), from the
anisotropy parameters $M_{a}$, which characterize the model's anisotropy;
now the quantities $\ln g=\sum_{a}q^{a}-2n\ln R_{0}$ and $M_{a}$ can be
expressed in terms of the new set of Misner-Chitr\`{e}-like variables $\tau $
and $y^{j}$ ($j=1,2...(n-1)$), as follows: 
\begin{equation}
\ln g=-ne^{-\tau }\frac{1+y^{2}}{1-y^{2}}\,\quad M_{a}\left( y^{j}\right) =%
\frac{1}{n}\left( 1+\frac{2y^{j}A^{a}{}_{i}}{1+y^{2}}\right) \;,  \label{AP}
\end{equation}
where $A^{a}{}_{i}$ is the matrix (\ref{matrA}). The Misner variables ${
\beta }^{j}$ are related to $y^{i}$ by 
$$
{\beta }^{j}=-e^{-\tau }2y^{j}/\left( 1-y^{2}\right) ,\,\,\, \alpha =\ln R_{0}^{2}-e^{-\tau }\left( 1+y^{2}\right) /\left( 1-y^{2}\right) \, .
$$ 
The parameterization (\ref{AP}) is
defined within the domain $-\infty <\tau <\infty $, $0<y<1$ ($y\equiv \sqrt{%
\sum_{j}(y^{j})^{2}}$) and (with $0\leq g\leq 1$) an appropriate choice of
the function $R_{0}$ allows one to cover all of the classically allowed region
of the configuration space using this parameterization.

Within this choice of variables, the evolution of the scale functions is
described by the action 
\begin{equation}
I=\int \left\{ \left( {\vec{P}_{\vec{y}}}\frac{\partial \vec{y}}{\partial t}%
+h{\frac{\partial \tau }{\partial t}}+P_{n}{\frac{\partial {\beta }^{n}}{%
\partial t}}\right) -\frac{Ne^{2\tau }}{n\left( n-1\right) R_{0}^{n}\sqrt{g}}%
\left[ \varepsilon ^{2}-h^{2}+\Pi +e^{-2\tau }P_{n}^{2}\right] \right\}
d^{n}xdt,  \label{Act}
\end{equation}
where $\vec{P}_{\vec{y}}$ and $h$ denote respectively the conjugate momenta
to $\vec{y}$ and $\tau $, while ${\varepsilon }^{2}=\frac{1}{4}\left(
1-y^{2}\right) ^{2}{\vec{P}_{\vec{y}}}^{2}$ and the potential term $\Pi $
has the following structure 
\begin{equation}
\Pi =n\left( n-1\right) R_{0}^{2n-2}e^{-2\tau }\sum_{a,b,c}\lambda
_{abc}g^{1+M_{a}-M_{b}-M_{c}}+n\left( n-1\right) R_{0}^{2n}ge^{-2\tau }V({%
\beta }^{n}).  \label{V}
\end{equation}
Here the coefficients $\lambda _{abc}$ (constructed by the spatial
derivatives of the reference vectors $\ell _{\alpha }^{a}$) are slow
functions of $\ln g$, i.e. of the time variable, and characterize the initial
intensity of the inhomogeneity field. When $g\ll 1$, we can use the
approximation of deep oscillations \cite{BKL70,BKL82} \footnote{%
It is a well-known result that, close enough to the singularity, the
inhomogeneous Mixmaster evolution can be represented as a sequence of Kasner
epochs 
(i.e. completely neglecting the potential term), 
while the
transition from one epoch to the next one can be regarded as an instantaneous
phenomenon (i.e. instant bounces against the potential walls). This
well-established oscillating behavior is called the ``deep oscillation
approximation'' and corresponds in our Hamiltonian picture to the
replacement of the actual potential term with a set of infinite potential
walls.}, in which the above potential is modeled by a set of potential
walls, each of them having the form
\begin{equation}
g^{\sigma _{a}}\rightarrow \theta _{\infty }[\sigma _{a}]=\left\{ 
\begin{array}{ll}
+\infty \,\,,\,\, & \sigma _{a}<0, \\ 
0\,\,\,,\qquad & \sigma _{a}>0,
\end{array}
\right.  \label{Apr}
\end{equation}
As a result the whole potential becomes asymptotically ($\Pi \rightarrow \Pi
_{\infty }$) independent of Kasner vectors, i.e. $\Pi _{\infty }=\sum \theta
_{\infty }\left( \sigma _{a}\right) $.

By solving the Hamiltonian constraint $H^0=0$ in (\ref{Act}) we define the
ADM action \cite{ADM}, reduced to the physical phase space, by a standard
procedure which leads to the following (reduced) action
\begin{equation}
I_{red}=\int ({\vec{P}}_{\vec{y}}\cdot \frac{d\vec{y}}{d\tau }+Q{\frac{%
\partial }{\partial \tau }}q-H_{ADM})d^{n}xd\tau ,  \label{5b}
\end{equation}
where
\begin{equation}
H_{ADM}\equiv -h=\sqrt{\varepsilon ^{2}+e^{-2\tau }Q^{2}+\Pi }
\end{equation}
is the ADM Hamiltonian and $\tau $ now plays the role of the time variable ($%
\dot{\tau}=1$), once the gauge is fixed by $N_{ADM}=\frac{n\left( n-1\right)
R_{0}^{n}\sqrt{g}}{2H_{ADM}}e^{-2\tau }$. For convenience 
we have
also redefined the scalar field variables as follows $q={\beta }^{n}$ and $%
Q=P_{n}$.

Thus, in the limit $\tau \rightarrow -\infty $, which corresponds to $\Pi
\rightarrow \Pi _{\infty }\left( y^{j}\right) $ (the independence of the
potential walls on the time variable $\tau $ is the most relevant advantage
in using Misner-Chitr\`{e}-like variables), the system (\ref{5b}) is nothing
more than a point-like realization of a billiard on the $\left( n-1\right) -$%
dimensional Lobachevsky space. By other words, in each point of the space,
the system dynamics is isomorphic to the motion of a point particle on the
(negative constant curvature) ($n-1$)-dimensional hypersurface formed by all
the admissible values $y^{j}$ (the real gravitational degrees of freedom). On
this domain the potential walls cut a region, which in dimensions $n\leq 9$
has a finite volume\footnote{
In the presence of dilaton-$p$-form fields billiards have finite volumes in
all dimensions, e.g., see Refs. \cite{DH,DH1}.} and therefore (in the absence
of a scalar field) the resulting billiard exhibits strong mixing properties 
\cite{KM95}. The role of scalar field (whose momentum, in this
approximation, is simply an arbitrary ``constant'' function $Q\left(
x^{i}\right) $) in the evolution of metric functions consists of making
geodesic lines on the billiard to be of a finite length and therefore of
suppressing the chaotic regime (no other bounces against the potential walls
take place).

Thus, with this scheme in our hands, we are now able to easily specify the
appropriate inequalities characterizing different dynamical regimes. When
approaching the singularity, the chaoticity can develop only in those
regions of the universe where the energy of the scalar field is sufficiently
small $e^{-2\tau }Q^{2}\ll \varepsilon ^{2}$. We also observe that the
condition for applying the approximation (\ref{Apr}) corresponds to free
motion in the allowed domain and therefore it can be written as follows
\begin{equation}
\varepsilon ^{2}\gg U.  \label{in}
\end{equation}

From the condition that the approximation of deep oscillations (\ref{Apr})
breaks at the moment $g\sim 1$ ($\tau \sim 1$), we find that the 
function $R_{0}$ should be
chosen as follows $R_{0}^{2n-2}=\frac{\varepsilon ^{2}}{n\left( n-1\right)
\lambda ^{2}}e^{2\tau }$ (where $\lambda ^{2}=\left| \sum  \lambda_{abc}\right|
$), so that in fact the inequality (\ref{in}) then just becomes
$g\ll 1$ ($\tau \ll 1$).

\section{Dependence of the Universe Evolution on the Initial Conditions}

In this section we develop a synthesis of the interpolating regime 
constructed above, but in view of the
different dynamical issues can take the
universe in consequence of different initial conditions. 
All the
considerations presented in the discussion below can be easily derived from
the Hamiltonian function and equations associated with the reduced action $%
I_{red}$.

As emphasized above \cite{BK73}, in the presence of a scalar field, the
final stage of the cosmological collapse ends with a monotonic Kasner-like
behavior, i.e. the number of oscillations is always finite. From a
phenomenological standpoint, if we consider an isotropic universe, this
means that the effective ``equation'' of state for the gravitational waves
(or the anisotropy parameters) is slightly softer then
the one for the scalar field
(behaving like stiff matter, characterized by $\varepsilon =p$), i.e., $%
\varepsilon \gtrsim p$. To see this explicitly, we may use some of the
results derived above.

Indeed near the singularity we have $\Pi \rightarrow \Pi _{\infty }\left(
y\right) $ and the scalar field momentum does not depend explicitly on the
time variable $Q=const$ and therefore the same behavior characterizes the
ADM ``energy'' of the anisotropy $\epsilon _{h}^{2}=\epsilon ^{2}(\vec{y},%
\vec{P}_{\vec{y}})=const$ (clearly from its expression, this quantity does
not depend explicitly on $\tau $). However, the ADM energy of the scalar
field depends on time $\epsilon _{q}^{2}=(Q)^{2}e^{-2\tau }$ and we get the
relation $\epsilon _{q}/\epsilon _{h}\sim e^{-\tau }\sim \ln g$. Thus in
the limit $\tau \rightarrow -\infty $ ($g\rightarrow 0$) the scalar field
dominates and the ADM Hamiltonian does not depend on the gravitational
variables $H_{ADM}\rightarrow \epsilon _{q}$, i.e. it turns out $\vec{y}%
=const$, $\vec{P_{\vec{y}}}=const$ (this corresponds to a stable Kasner
regime approaching the singularity).

On the other hand, during the expansion of the universe the role played by the
scalar field through the evolution becomes smaller and smaller an,
therefore the following various types of dynamical regimes can take place.

\subsection{The vacuum-type regime}

If the initial ADM energy of the scalar field is not very big,
then the
ADM energy of the anisotropy starts to dominate and this
type of regime will
be described by the oscillatory behavior with a frozen scalar field $Q=const$,
$q=const$. The scalar field potential, which is in general
always negligible near the singularity
(i.e. at sufficiently high
temperatures), in this case behaves like an effective 
cosmological constant $V\left( q\right)=const$.
However if this constant remains small
as compared to the ADM energy of the anisotropy
during the whole period of
applicability of the BKL picture,
then this would result in the vacuum-type evolution.

To define the moment $\tau_in$ when the BKL approximation breaks down
(i.e., $g\sim 1$), we 
consider the synchronous cosmological time, which is related to $\tau $ 
by the equation 
\begin{equation}
dt= N_{ADM}d\tau=\frac{n\left( n-1\right)
R_{0}^{n}\sqrt{g}}{2H_{ADM}}e^{-2\tau }d\tau .
\end{equation}
In the vacuum stage we have $H_{ADM}=\epsilon _{h}=const$,
which gives us
\begin{equation}\label{tt}
\sqrt{g} \sim  \frac{t/t_{in}}{1-\ln (t/t_{in})} ,
\end{equation}
where $t_{in} =\frac{2(n-1)R_0^n}{n\epsilon _h}$ and $R_{0}^{2n-2}=\frac{\varepsilon ^{2}}{n\left( n-1\right)
\lambda ^{2}}e^{2\tau }$.
We recall that during the expansion,
 $R_0$ remains a constant in the oscillatory BKL regim,
since as shown in Ref.~\cite{K93,KM95}
$\lambda \sim \lambda _0 e^{\tau }$.
 
From the physical standpoint the moment $\tau _{in}$
($g_{in}\sim 1$) corresponds to the case
where the characteristic scale of the
inhomogeneity $L_{in}$ matches the horizon size $L_{h}$, $L_{i}\sim L_{h}$,
the so-called moment of origin of a stable background e.g., see \cite
{KMq97}.

\subsection{The inflationary-type regime}

This kind of regime is realized when the scalar field
potential grows until it is
comparable with the ADM energy of anisotropy
before the end of the BKL oscillatory regime.
This dynamical feature results in the so-called
inflationary-like evolution, replacing the last
Kasner epoch of the oscillatory regime
(recall
that if the scalar field potential remains small, then
the evolution corresponds
to the vacuum case).

Under the assumption that the potential term
$gV\left( q\right) $ starts to
dominate before the end of the Mixmaster
$\tau <\tau _{in}\sim 1$, then the ADM energy can be approximated by a
function of the form
\begin{equation}
H_{ADM}\approx \sqrt{\epsilon _{h}^{2}+n\left( n-1\right) R_0^{2n}ge^{-2\tau
}V\left( q\right) }  \label{i}
\end{equation}
(where we neglect the kinetic term of the scalar field
and the spatial
curvature term also). Thus the conditions for this regime
to exit are expressed via the
inequalities
\begin{equation}
\epsilon _{q}^{2}\ll \epsilon _{h}^{2}\sim n\left( n-1\right)
R_0^{2n}g_{c}e^{-2\tau _{c}}V\left( q\right)\ ,   \label{sin}
\end{equation}
where $\tau _{c}$ corresponds to the beginning
of the inflationary
evolution, i.e. the moment at which both terms
at right hand side of (\ref
{sin}) have the same order, that is to say 
$$
\epsilon _{h}^{2}\sim \ n\left( n-1\right)
R_{0}^{2n}g_{c}e^{-2\tau _{c}}V( q) 
$$ 
and the inequality which
expresses the applicability of the Mixmaster approximation reads $g_{c}\ll
g_{in}\sim 1$, or equivalently,
in terms of the synchronous cosmological time,
as follows from (\ref{tt})
\begin{equation}
t_c \ll t _{in}\;\, , 
\end{equation}
since $t_c$ is given by
\begin{equation}
t_c = \sqrt{\frac{n-1}{nV(q)}}. 
\end{equation}

From a physical point of view the last inequality
expresses the well known
condition $L_{i}\gg L_{C}$ for the inflationary dynamics, where $L_{C}$
is
related to the field ``Compton scale''. 
We also recall that at later times, i.e., as $t>t_c$,
the inflationary evolution
leads to the scalar curvature term in
$H_{ADM}$ overtaking the
kinetic energy of the anisotropy $\epsilon ^2 _{h}$.
However, we emphasize that both these terms
will remain very small compared to the
effective cosmological constant $V(q)$.

We note that the inflationary regime has a finite duration,
due to a `slow'
evolution of the scalar field
(the so-called {\it slow-rolling phase}),
which is well described in the canonical literature on inflation \cite
{KT90}. The slow evolution phase of the
scalar field can be found from the
Hamiltonian equations
\begin{equation}
{\frac{\partial }{\partial \tau }}q=\frac{Qe^{-2\tau }} {H_{ADM}}\approx 0,
\end{equation}
and
\begin{equation}
{\frac{\partial }{\partial \tau }}Q=-\frac{1}{2}\frac{n\left( n-1\right)
ge^{-2\tau }R_0^{2n}V\left( q\right) ^{\prime }}{H_{ADM}}.
\end{equation}
Hence in the synchronous gauge ($N=1$) $dt/d\tau ={\frac{n(n-1)}{2}R_0}^{n}
\sqrt{g}\exp (-2\tau )/H_{ADM}$ we get
\begin{equation}
{\frac{\partial }{\partial t}}q=\frac{2Q}{n(n-1)R_0^{n}\sqrt{g}},
\end{equation}
\begin{equation}
{\frac{\partial }{\partial t}}Q=-R_0^{n}\sqrt{g}V\left( q\right) ^{\prime },
\end{equation}
where the initial conditions should be chosen so that
\begin{equation}
Q_{0}^{2}\ll \epsilon _{h}^{2}e^{2\tau _c}\lesssim n\left( n-1\right)
R_0^{2n}g_{c}V\left( q_{c}\right) \;.
\end{equation}
The rate of expansion is described by the equation
$dt/d\tau $ which can be
rewritten as follows (we take $\ln g=-ne^{-\tau }${\ (}$y=0$) and $H_{ADM}=
\sqrt{n\left( n-1\right) R_0^{2n}ge^{-2\tau }V\left( q\right) }$)
\begin{equation}
{\frac{\partial }{\partial t}}\sqrt{g}= \sqrt{{\frac{nV\left( q\right) }{
(n-1) }}}\sqrt{g},\;\sqrt{g}\sim \sqrt{g_{c}}\exp \left( \int_{t_c}^{t} 
\sqrt{{\ \frac{nV\left( q\right) }{(n-1)}}}dt\right) .
\end{equation}

\subsection{The scalar-field-dominated-type regime}

Finally we consider another type of regime which takes place when
the scalar field dominates during the entire Mixmaster
approximation, i.e. $\epsilon_{q}^{2}\gg \epsilon _{h}^{2}$.
In this case the solution has a monotonic
behavior; we can neglect the anisotropy functions
from the very beginning
and the evolution proceeds in the same way as in the case
of isotropic models in the
presence of a scalar field.

\subsection{Brief concluding remarks}

We conclude this section by emphasizing some
relevant features of the interpolating solution
we have obtained 
which help in getting physical insight into its cosmological
setting. 

i) It is remarkable that the condition ensuring
the inflationary scenario
starts well inside the range of validity of the oscillatory regime
is nothing more than the assumption at that moment $\tau _c$ that the
``Compton length'' associated with the scalar field
$L_{c}$ be much smaller than the inhomogeneous scale $L_{in}$ of the
universe; this condition is telling us that, 
as in the homogeneous case 
(when the condition for the existence of an inflationary stage
reads in this same form),
even in this general context the inflation can
start only if the spatial gradients are sufficiently small.
Thus this requirement
for the validity of the BKL approach 
which is at the heart of our
interpolating approach is just the inhomogeneous
generalization of the well-know standard restriction for
homogeneous inflation;
this fact provides very strong physical support
for the dynamical analysis developed here.

However, we must mention that the condition so obtained
is a point-like one and in principle, 
cannot take place in
some spatial domain. It is worth noting that
since we require that inflation starts before 
the BKL regime ends, i.e. when the horizon size is still much less
then the inhomogeneous scale ($L_{h}\ll L_{i}$),
the dynamical regime derived above can (nevertheless)
occur naturally in causal regions.

ii) An important feature of the inflationary scenario
consists of the frozen dynamics acquired by the
anisotropy of the universe.
In other words,
during this stage of the evolution,
the functions $\beta _+$ and $\beta _-$ remain almost constant.
As a consequence, the only evolving variable is now $alpha $,
i.e. the volume of the universe; 
this fact makes it evident that the dominant term
during the inflation is $\Lambda e^{3\alpha }$ and
any other term in the spatial curvature,
although increasing like (at most) $e^{2\alpha }$,
becomes more and more negligible.

iii) Finally we emphasize that the results obtained above do not
depend on the
particular form taken by the scalar field potential, as long as it realizes a
spontaneous symmetry breaking process and a sufficiently long phase of
slow-rolling \cite{CW73}.

It is also worth noting that the compatibility of the inflationary
scenario with our actual cosmic phenomenology seems to be confirmed by
recent observations on the microwaves background radiation \cite{dB01}, thus
allowing, on the basis of the analysis developed here, very general dynamical
behavior for the primordial phases of the universe evolution.

\vspace{0.5cm}

The referee is thanked for valuable
contributions in improving the form of
this paper.

\end{document}